\documentclass[twocolumn,english,aps,pra,showpacs]{revtex4}
\usepackage[T1]{fontenc}
\usepackage[latin9]{inputenc}
\usepackage{color}
\usepackage{babel}

\usepackage{amstext}
\usepackage{graphicx}
\usepackage[unicode=true, pdfusetitle,
 bookmarks=true,bookmarksnumbered=false,bookmarksopen=false,
 breaklinks=false,pdfborder={0 0 1},backref=false,colorlinks=true]
 {hyperref}
\hypersetup{
 citecolor=blue,linkcolor=blue}

\makeatletter

\providecommand{\LyX}{L\kern-.1667em\lower.25em\hbox{Y}\kern-.125emX\@}

\@ifundefined{textcolor}{}
{%
 \definecolor{BLACK}{gray}{0}
 \definecolor{WHITE}{gray}{1}
 \definecolor{RED}{rgb}{1,0,0}
 \definecolor{GREEN}{rgb}{0,1,0}
 \definecolor{BLUE}{rgb}{0,0,1}
 \definecolor{CYAN}{cmyk}{1,0,0,0}
 \definecolor{MAGENTA}{cmyk}{0,1,0,0}
 \definecolor{YELLOW}{cmyk}{0,0,1,0}
 }

\makeatother

\begin{document}

\title{Quantum correlations in topological quantum phase transitions}

\author{Yi-Xin Chen}

\email{yxchen@zimp.edu.cn}

\affiliation{Zhejiang Institute of Modern Physics, Zhejiang University, Hangzhou
310027, China}

\author{Sheng-Wen Li}

\affiliation{Zhejiang Institute of Modern Physics, Zhejiang University, Hangzhou
310027, China}

\begin{abstract}
We study the quantum correlations in a 2D system that possesses a
topological quantum phase transition. The quantumness of two-body
correlations is measured by quantum discord. We calculate both the
correlation of two local spins and that of an arbitrary spin with
the rest of the lattice. It is notable that local spins are classically
correlated, while the quantum correlation is hidden in the global
lattice. This is different from other systems which are not topologically
orderd. Moreover, the mutual information and global quantum discord
show critical behavior in the topological quantum phase transition.
\end{abstract}
\maketitle

\section{Introduction}

Topological phase is a new kind of order that cannot be described
by the symmetry-breaking theory \cite{wen_quantum_2004}. A typical
example is the quantum Hall system, which exhibits a lot of amazing
properties, such as topological degeneracy and fractional statistical
behaviors . Especially, the property of topological protection may
lead a new way for quantum computation \cite{nayak_non-abelian_2008}.

Different from the quantum Hall system, Kitaev toric code model is
an exactly solvable spin lattice model that is topologically ordered
\cite{kitaev_fault-tolerant_2003}. The system is immune to small
perturbations. The breaking down of the topological phase happens
through a quantum phase transition \cite{s._sachdev_quantum_1999}.
A lot has been studied about the topological quantum phase transition,
especially about the toric code model in the present of a magnetic
field \cite{trebst_breakdown_2007,vidal_self-duality_2009}.

Concepts of quantum information are borrowed to the study of quantum
phase transition, like entanglement and fidelity \cite{osterloh_scaling_2002,ma_many-body_2009}.
Here, we are interested in the correlations in topological phase ,
because the magic power of quantum computation roots from the strange
non-classical correlations.

Entanglement is the most important non-classical correlation in quantum
information processing, such as quantum teleportation \cite{bennett_teleportingunknown_1993}.
However, some separable states also have properties that are not achievable
by classical methods \cite{ollivier_quantum_2001}. Recent results
suggest that these correlations may also take effect in quantum computation.
The classification of non-classical correlations and their effects
still remains an open problem \cite{modi_relative_2009}.

Quantum discord is a measurement for the ``quantumness'' of a pairwise
correlation \cite{ollivier_quantum_2001}. It is based on the fact
that the mutual information has two equivalent definitions in the
classical world, while their quantum generations are not equivalent.
Quantum discord is defined as the minimum of their difference and
measures how ``quantum'' the correlation is. Besides entangled states,
some separable states also have non-zero quantum discord, which means
they are non-classical. Recent studies suggest that quantum discord
but not entanglement may be responsible for mixed-state computation
\cite{datta_quantum_2008,lanyon_experimental_2008}.

The quantum discord in quantum phase transition were studied in 1D
systems, such as $XXZ$-chain and some other $Z_{2}$-symmetric 1D
spin models \cite{dillenschneider_quantum_2008,sarandy_classical_2009}.
During the phase transition, quantum discord shows critical behavior
at the phase transition point. The study of quantum discord in thermal
Heisenberg system also shows some different behavior from entanglement
\cite{werlang_thermal_2009}.

In this paper, we study the correlations in Castelnovo-Chamon model
\cite{castelnovo_quantum_2008}, which is a 2D system that shows a
quantum phase transition from a topologically ordered phase to a magnetized
one. It is a deformation of toric code model and possesses higher
symmetry than other 1D models mentioned above. Both local spin-spin
correlation and that between a spin and the rest of the whole lattice
are calculated. It is notable that in such a topologically ordered
system, quantum discord of local spins is always zero in both phases,
which means the local correlations are totally classical. While the
correlation between a local spin and the rest of the lattice behaves
more like a pairwise entangled pure state, and the quantum discord
signals critical point. This is different from previous studies in
other models \cite{dillenschneider_quantum_2008,sarandy_classical_2009,werlang_thermal_2009}.
Our results shows that in topologically ordered system, the quantum
correlation is hidden in the lattice globally by the high symmetry
of the system.

Besides, we calculated the mutual information of the correlations
in the system. It was pointed out that in topologically ordered systems,
which have no local order parameter, the topological quantum phase
transition can be signaled by local properties like the reduced fidelity
of two spins, and it is even more sensitive than the global fidelity
\cite{eriksson_reduced_2009,abasto_fidelity_2008}. Here, we calculate
the mutual information, both of the global correlation and that of
two local spins. We see that the mutual information could also characterize
the critical behavior.

The paper is organized as follows. In Sec. II we briefly review the
concept of quantum discord and the basic properties. In Sec. III,
we introduce the Castelnovo-Chamon model. We give the ground state
and explain how it can be mapped to the classical Ising model. In
Sec. IV, we calculate the quantum discord of local spin-spin correlations,
and also that of a local spin with the rest of the whole lattice.
Conclusion is drawn in Sec. V.

\section{Quantum Discord}

Quantum discord can be used as a measure of the quantumness of a pairwise
correlation \cite{ollivier_quantum_2001}. In this part, we briefly
introduce this concept.

Information is amount of uncertainty that can be eliminated after
we get a measurement result. Mutual information $\mathcal{I}(A:B)$
describe the information about $A$ we gain after the measurement
of $B$, or rather, the information that $A$ and $B$ have in common
\cite{nielsen_quantum_2000}. In classical world, there are two equivalent
definitions of mutual information,\begin{eqnarray*}
\mathcal{I}(A:B) & = & H(A)+H(B)-H(A,B),\\
\mathcal{J}(A:B) & = & H(A)-H(A|B),\end{eqnarray*}
with the Shannon entropy $H(\cdot)=-\sum_{i}p_{i}\log_{2}p_{i}$.
$H(A|B)$ is the conditional entropy, which is a measure of how uncertain
we are about $A$, averagely, when $B$ is known. The Bayes' law tells
us that $p(a_{i},b_{j})=p(a_{i}|b_{j})p(b_{j})=p(b_{j}|a_{i})p(a_{i})$,
where $p(a_{i}|b_{j})$ is the conditional probability that describes
the probability to get $a_{i}$ when we know the value of $B$ is
$b_{j}$. That guarantees the equivalence of the two definitions above.

However, things are different when we generalize the concepts to the
quantum world. We can get the quantum version of $\mathcal{I}(A:B)$
easily by replacing $H$ with von Neumann entropy $S(\rho)=-\textrm{tr}(\rho\log\rho)$.
While the concept of conditional entropy in fact implicitly calls
for a measurement of $B$. To get the corresponding $\mathcal{J}(A:B)$,
we have to choose a set of basis $\{\hat{\Pi}_{i}^{B}\}$ to measure
system $B$. The state of the system after measurement is $\rho_{i}=\hat{\Pi}_{i}^{B}\rho_{AB}\hat{\Pi}_{i}^{B}/p_{i}$,
where $p_{i}=\textrm{tr}(\hat{\Pi}_{i}^{B}\rho_{AB}\hat{\Pi}_{i}^{B})$.
With the knowledge we gain after the measurement, we get\begin{eqnarray}
\mathcal{J}(A|\{\hat{\Pi}_{i}^{B}\}) & = & S(\rho_{A})-S(\rho_{AB}|\{\hat{\Pi}_{i}^{B}\})\nonumber \\
 & = & S(\rho_{A})-\sum_{i}p_{i}S(\rho_{i}).\end{eqnarray}
The value of $\mathcal{J}(A|\{\hat{\Pi}_{i}^{B}\})$ depends on the
choice of $\{\hat{\Pi}_{i}^{B}\}$, i.e., how we measure $B$, and
therefore may not be equal to $\mathcal{I}(A:B)$ any more. Quantum
discord is defined as the minimum of their difference,\begin{equation}
D(\rho)=\textrm{min}\left[\mathcal{I}(A:B)-\mathcal{J}(A|\{\hat{\Pi}_{i}^{B}\})\right].\end{equation}

$\mathcal{J}(A|\{\hat{\Pi}_{i}^{B}\})$ describes the amount of information
of $A$ achievable by projective measurements on $B$. It can be proved
that $D(\rho)\geq0$. From the derivation above, we see that the quantum
discord of classical correlations should be zero. We can use quantum
discord as a measure of the {}``quantumness'' of a two-body correlation.

The quantum discord of states that contain entanglement is obviously
non-zero. It should be emphasized again that not all separable states
are classical under the definition of quantum discord. A simple example
is $\rho=\left(|00\rangle\langle00|+|++\rangle\langle++|\right)/2$,
where $|+\rangle=(|0\rangle+|1\rangle)/\sqrt{2}$. The quantum discord
is non-zero. The information inside the state cannot be fully extract
just by local projective measurements. From the description above
we can see that the quantum discord of $\rho_{AB}$ is zero, if and
only if $\rho_{AB}$ has the form of $\rho_{AB}\sim\sum p_{i}\rho_{i}^{A}\otimes\Pi_{i}^{B}$.

\section{Castelnovo-Chamon Model}

Castelnovo and Chamon proposed a model that shows topological quantum
phase transition \cite{castelnovo_quantum_2008}. It is a deformation
of Kitaev toric code model. The Hamiltonian is

\begin{equation}
H=-\lambda_{0}\sum_{p}B_{p}-\lambda_{1}\sum_{s}A_{s}+\lambda_{1}\sum_{s}\exp\left(-\beta\sum_{i\in s}\hat{\sigma}_{i}^{z}\right),\label{eq-H}\end{equation}
where $\lambda_{0,1}>0$, $A_{s}=\prod_{i\in s}\hat{\sigma}_{i}^{x}$
and $B_{p}=\prod_{i\in p}\hat{\sigma}_{i}^{z}$ are the \emph{star}
and \emph{plaquette operators} in toric code model respectively. $\beta$
is a coupling constant. The star operator $A_{s}$ acts on the four
spins around the vertex $s$, while the plaquette operator $B_{p}$
acts on the four spins on the edges of the plaquette $q$, as shown
in Fig. 1. We consider the problem under the torus boundary condition.

\begin{figure}
\includegraphics[width=6cm]{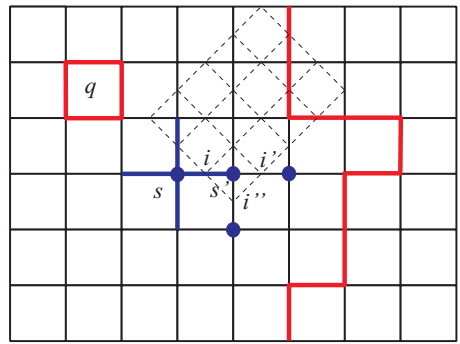}\caption{(Color online). Demonstration of the model. Spins lie on the edges
(like $i,i',i'')$. The nearest spins $i,i'$ become next-nearest
in the dual lattice (the dashed line). The red plaquette and the blue
cross represent the plaquette and star operators. The red line across
the lattices represents a product of $\sigma_{i}^{z}$ along the non-trivial
loop on a torus. The system is invariant under this transformation.}

\end{figure}

The ground state can be written down analytically. We give the state
in the topological sector that contains the fully magnetized state
$|0\rangle=|\uparrow\uparrow\uparrow\cdots\uparrow\rangle$ as\begin{equation}
|\textrm{GS}(\beta)\rangle=Z(\beta)^{-\frac{1}{2}}\sum_{g\in G}\exp\left[\beta\sum_{i}\sigma_{i}^{z}(g)/2\right]g|0\rangle,\label{eq-gs}\end{equation}
with $Z(\beta)=\sum_{g\in G}\exp\left[\beta\sum_{i}\sigma_{i}^{z}(g)\right]$.
$G$ is the Abelian group generated by the star operators $\left\{ A_{s}\right\} $.
So $g|0\rangle$ contains separable spins taking the form like $|011110\dots0\rangle$,
and we can denote each $g|0\rangle$ by a corresponding binary number
as $|x\rangle$ (as what we do in the following). $\sigma_{i}^{z}(g)$
is the value of spin at site $i$ in state $g|0\rangle$. The sum
in the exponential term in fact counts the total magnetic polarization
of $g|0\rangle$.

It may be not obvious to get Eq. (\ref{eq-gs}) directly. We can just
put it back into Eq. (\ref{eq-H}) and it can be easily checked. When
$\beta=0$ the model reduces to the toric code model. When $\beta\rightarrow\infty$
the ground state becomes the fully magnetized reference state $|0\rangle.$
At $\beta_{c}=(1/2)\ln(\sqrt{2}+1)$ there is a second-order topological
quantum phase transition, according to the study of topological entropy
\cite{castelnovo_quantum_2008} and fidelity \cite{abasto_fidelity_2008,eriksson_reduced_2009}
in this model.

Furthermore, the value of $\sigma_{i}^{z}(g)$ (note that there is
no hat, it is just an integer number relating to $g$) of the $i$th
spin is actually determined by whether the two ends ($s$ and $s'$
in Fig. 1) of the $i$th edge are acted by $A_{s(s')}$ or not. As
$A_{s}^{2}=\mathbf{1}$, elements of $G$ can be represented as a
configuration of $\{\theta_{s}\}$, where $\theta_{s}=+1$ means $A_{s}$
acts on vertex $s$, while $\theta_{s}=-1$ means not. So we get $\sigma_{i}^{z}(g)=\theta_{s}\theta_{s'}$.
The normalizer in the ground state Eq. (\ref{eq-gs}) is\begin{equation}
Z(\beta)=\sum_{\{\theta_{s}\}}\exp\left[\beta\sum_{\langle ss'\rangle}\theta_{s}\theta_{s'}\right],\end{equation}
which is just the canonical partition function of 2D classical Ising
model without external field, with the Hamiltonian $H_{\textrm{Ising}}=-\sum_{\langle ss'\rangle}\theta_{s}\theta_{s'}$.

As an example, we can calculate the correlation function as\begin{eqnarray}
\langle\textrm{GS}|\hat{\sigma}_{i}^{z}|\textrm{GS}\rangle & = & \sum_{g\in G}\sigma_{i}^{z}(g)\exp\left[\beta\sum_{j}\sigma_{j}^{z}(g)\right]\nonumber \\
 & = & \sum_{\{\theta_{s}\}}\theta_{s}\theta_{s'}\exp\left[\beta\sum_{\langle ss'\rangle}\theta_{s}\theta_{s'}\right]\\
 & = & \langle\theta_{0,0}\theta_{0,1}\rangle_{\textrm{Ising}}.\nonumber \end{eqnarray}
Thus, we can see that the model can be mapped to the Ising model,
which is exactly solvable \cite{mccoy_two-dimensional_1973}.

\section{Correlations in the Lattice}

In this section, we discuss the correlations in the lattices. Both
the local spin-spin correlation and the global correlation between
a single spin with the rest of the lattice are considered. We find
that the local correlations are classical, and the quantum correlation
emerges only when considering the whole lattice. In both cases, the
mutual information signals the critical behavior.

\subsection{Local spin-spin correlation}

Firstly, let us look at the correlation of two local spins. We need
to get the reduced density matrix of two spins. The set $\left\{ \frac{1}{2}\hat{\sigma}_{i}^{\mu}\hat{\sigma}_{j}^{\nu}\right\} $,
where $\mu,\nu$ take value $0,1,2,3$ and $\hat{\sigma}^{0}=\mathbf{1}$,
contains 16 matrices and they form a complete orthonormal basis for
$4\times4$ Hermitian matrices under the Hilbert-Schmidt inner product
$(A,B)_{\textrm{H-S}}\equiv\textrm{tr}(A^{\dagger}B)$ \cite{nielsen_quantum_2000}.
Conveniently, the reduced density matrix can be wirtten as the expansion
of the basis set $\left\{ \frac{1}{2}\hat{\sigma}_{i}^{\mu}\hat{\sigma}_{j}^{\nu}\right\} $
\cite{wang_pairwise_2002,eriksson_reduced_2009},\begin{equation}
\hat{\rho}_{ij}=\frac{1}{4}\sum_{\mu,\nu=0}^{3}\langle\hat{\sigma}_{i}^{\mu}\hat{\sigma}_{j}^{\nu}\rangle\hat{\sigma}_{i}^{\mu}\hat{\sigma}_{j}^{\nu},\label{eq-expand}\end{equation}
where $\langle\hat{\sigma}_{i}^{\mu}\hat{\sigma}_{j}^{\nu}\rangle=\textrm{Tr}(\hat{\rho}_{\textrm{GS}}\hat{\sigma}_{i}^{\mu}\hat{\sigma}_{j}^{\nu})=\textrm{tr}(\hat{\rho}_{ij}\hat{\sigma}_{i}^{\mu}\hat{\sigma}_{j}^{\nu})$
is the inner product of $\hat{\rho}_{ij}$ and $\hat{\sigma}_{i}^{\mu}\hat{\sigma}_{j}^{\nu}$.

\begin{figure}
\includegraphics[width=7cm]{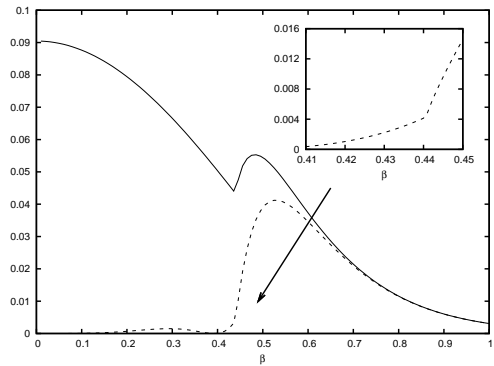}

\caption{Mutual information of two nearest spins $\sigma_{i}^{z}\sigma_{i''}^{z}$
(the solid line) and $\sigma_{i}^{z}\sigma_{i'}^{z}$ (the dashed
line) as shown in Fig. 1. Critical change happens at $\beta_{c}=(1/2)\ln(\sqrt{2}+1)$,
which is in accord with previous studies. The quantum discord of the
two spins are always zero. }

\end{figure}

Furthermore, most terms above can be eliminated because of the symmetry
of the system. Draw a closed loop through the torus arbitrarily (as
the red line shown in Fig. 1), and define a corresponding transformation
$\hat{\mathcal{P}}=\prod_{line}\hat{\sigma}_{i}^{z}$. The Hamiltonian
Eq. (\ref{eq-H}) is invariant under the transformation $\hat{\mathcal{P}}$.
Also, $\hat{\rho}_{ij}$ should commute with any $\hat{\mathcal{P}}$.
Only the terms $\mathbf{1},\hat{\sigma}_{i}^{z},\hat{\sigma}_{j}^{z}$
and $\hat{\sigma}_{i}^{z}\hat{\sigma}_{j}^{z}$ could exist, so we
get\begin{eqnarray}
\hat{\rho}_{ij} & = & \frac{1}{4}\left(\mathbf{1}+\langle\hat{\sigma}_{i}^{z}\rangle(\hat{\sigma}_{i}^{z}+\hat{\sigma}_{j}^{z})+\langle\hat{\sigma}_{i}^{z}\hat{\sigma}_{j}^{z}\rangle\hat{\sigma}_{i}^{z}\hat{\sigma}_{j}^{z}\right).\end{eqnarray}

The density matrix is diagonal. It can be written in the form of $\hat{\rho}_{ij}\sim\sum p_{n}\rho_{n}\otimes\Pi_{n}$.
According to what we have seen in Sec. II, the quantum discord of
$\hat{\rho}_{ij}$ is zero. That means the correlations between any
two local spins are always classical. This is quite different from
other studies of quantum discord in the phase transition of 1D systems
that are not topologically ordered \cite{dillenschneider_quantum_2008,sarandy_classical_2009,werlang_thermal_2009},
where the quantum discord of local spins shows different behavior
in different phase areas and exhibits critical behavior.

This roots from the high symmetry of the topologically ordered system.
This 2D system exhibits higher symmetry than other 1D $Z_{2}$-symmetric
models \cite{sarandy_classical_2009,dillenschneider_quantum_2008}.
The system is conserved under the transformation of $\hat{\mathcal{P}}$
along any closed loop, which eliminates all non-diagonal terms. So
the quantum discord is zero in the topological phase area. It was
stated in Ref. \cite{nayak_non-abelian_2008} that in a topologically
ordered system, all observable properties should be invarint under
smooth deformations (diffeomorphisms) of the space-time manifold,
which means the only local operator that has nonvanishing correlation
functions is the identity. For example, in toric code model, only
identity exists in the expansion Eq. (\ref{eq-expand}) and $\hat{\rho}_{ij}\sim\mathbf{1}\otimes\mathbf{1}$,
which means local spins are even uncorrelated. While the other phase
area where $\beta\rightarrow\infty$ is a fully magnetized phase,
which obviously only contains classical information of probability
. This is why local correlations are classical in both phases.

Besides, it was stated that topological quantum phase transition cannot
be described by the symmetry-breaking of a local order parameter and
involves a global rearrangement of non-local correlations \cite{wen_quantum_2004}.
However, recent researches indicated that some concepts in quantum
information theory, which describe local properties although, still
signal the singularity in topological quantum phase transition \cite{trebst_breakdown_2007,castelnovo_quantum_2008,eriksson_reduced_2009}.
The reduced fidelity and local magnetization were studied in the Castelnovo-Chamon
model and the Kitaev toric code in a magnetic field and they exhibit
critical behavior of the topological quantum phase transition. We
calculate the mutual information $\mathcal{I}$ of two nearest spins,
which is also a local property (Fig. 2). The correlations in $\hat{\rho}_{ij}$
can be evaluated with the help of the mapping to Ising model, as mentioned
in Sec. III. We can see that the mutual information of both nearest
and next-nearest spins (in the dual lattice) exhibits critical behavior.
But the next-nearest mutual information is much less sensitive.

\subsection{Global correlation in the lattice}

As we have seen in the last part, the correlations between local spins
are completely classical in both phases. In this part, we calculate
the correlation between a local spin and the rest of the whole lattice.
As the increasing of $\beta$, the system turns to the magnetic phase,
and the correlation between a local spin and the lattice becomes more
and more ``classical''.

To calculate the correlation of an arbitrary spin denoted by $k$
with the lattice, we treat the rest of the lattice as a whole system.
We can always rewrite the ground state as\begin{eqnarray}
|\textrm{GS}(\beta)\rangle & = & \sum_{x}a_{x}|x\rangle|0\rangle_{k}+\sum_{y}b_{y}|y\rangle|1\rangle_{k}\nonumber \\
 & = & a|X\rangle|0\rangle_{k}+b|Y\rangle|1\rangle_{k}.\label{eq-gs2}\end{eqnarray}
where $|X\rangle=\sum_{x}a_{x}|x\rangle$ and $|Y\rangle=\sum_{y}b_{y}|y\rangle$.
The $x,y$ in the basis vectors are the binary number representation
of $g|0\rangle$ excluding the $k$th spin, as mentioned in Sec. III.
Notice that $g|0\rangle$ and $g'|0\rangle(g\neq g')$ have at least
four different spins. So we are sure that $|X\rangle$ and $|Y\rangle$
have no term in common, and $\langle X|Y\rangle=0$. Therefore, we
can treat $|\textrm{GS}(\beta)\rangle$ as a simple $2\times2$ entangled
state. In this case, the quantum discord is equal to the entanglement
of entropy \cite{vedral_classical_2003,maziero_classical_2009},\begin{equation}
D(\rho_{AB})=\mathcal{I}(A:B)/2=S(A)=S(B).\end{equation}

We calculate it in detail. The coefficients $a_{x},b_{y}$ are superposition
coefficients in Eq. (\ref{eq-gs}) correspoindingly. So the value
of $a^{2}(b^{2})$ is just the Ising partition function with a constraint
that $\sigma_{k}^{z}=1(-1)$, in another word, $\theta_{r}\theta_{r'}=1(-1)$,
where $r$ and $r'$ are the nearest vertices of spin $k$.\begin{eqnarray}
a^{2} & = & \sum_{\{\theta_{s}\},\theta_{r}\theta_{r'}=1}\exp\left[\beta\sum_{\langle\theta_{s}\theta_{s'}\rangle}\theta_{s}\theta_{s'}\right]/Z(\beta),\nonumber \\
b^{2} & = & \sum_{\{\theta_{s}\},\theta_{r}\theta_{r'}=-1}\exp\left[\beta\sum_{\langle\theta_{s}\theta_{s'}\rangle}\theta_{s}\theta_{s'}\right]/Z(\beta).\end{eqnarray}

Notice that $a^{2}-b^{2}$ is just the nearest correlation function
$\langle\theta_{0,0}\theta_{0,1}\rangle$. \begin{eqnarray}
\langle\theta_{0,0}\theta_{0,1}\rangle & = & \left[\sum_{\theta_{r}\theta_{r'}=1}e^{\beta\sum\theta_{s}\theta_{s'}}-\sum_{\theta_{r}\theta_{r'}=-1}e^{\beta\sum\theta_{s}\theta_{s'}}\right]/Z(\beta)\nonumber \\
 & = & a^{2}-b^{2}.\end{eqnarray}
Together with $a^{2}+b^{2}=1$, we can get the value of $a,b$.

\begin{figure}
\includegraphics[width=7cm]{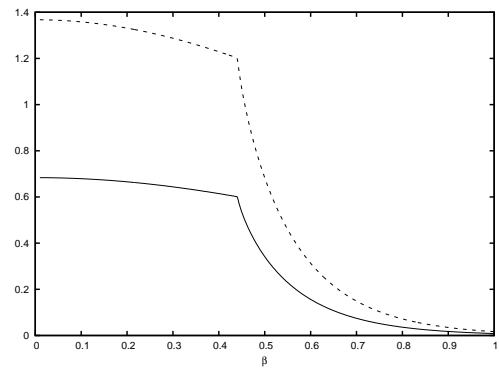}

\caption{Global correlations between a local spin and the rest of whole lattice.
Here, the quantum discord (the solid line) is equal to the entropy
of entanglement, and just one-half of the mutual information (the
dashed line) of the pairwise system. Both the quantum discord and
the mutual information show critical change at the phase transition
point.}

\end{figure}

Now we calculate the quantum discord of the ground state Eq. (\ref{eq-gs2}).
Instead of doing all the possible projective measurement to the spin,
equivalently, we implement all possible local unitary operations on
the spin and then measure it by $\{|0\rangle\langle0|,|1\rangle\langle1|\}$.

\begin{eqnarray*}
 &  & \Pi_{0}U^{\dagger}\hat{\rho}_{\textrm{GS}}U\Pi_{0}=\tilde{\rho}_{0}\otimes|0\rangle\langle0|\\
 & = & \left(\begin{array}{cc}
a^{2}\cos^{2}\frac{\theta}{2} & \frac{1}{2}ab\sin\theta e^{i\phi}\\
\frac{1}{2}ab\sin\theta e^{-i\phi} & b^{2}\sin^{2}\frac{\theta}{2}\end{array}\right)\otimes\Pi_{0},\\
 &  & \Pi_{1}U^{\dagger}\hat{\rho}_{\textrm{GS}}U\Pi_{1}=\tilde{\rho}_{1}\otimes|1\rangle\langle1|\\
 & = & \left(\begin{array}{cc}
a^{2}\sin^{2}\frac{\theta}{2} & -\frac{1}{2}ab\sin\theta e^{i\phi}\\
-\frac{1}{2}ab\sin\theta e^{-i\phi} & b^{2}\cos^{2}\frac{\theta}{2}\end{array}\right)\otimes\Pi_{1},\end{eqnarray*}
where \begin{equation}
U=\left(\begin{array}{cc}
\cos\frac{\theta}{2} & \sin\frac{\theta}{2}e^{-i\phi}\\
\sin\frac{\theta}{2}e^{i\phi} & \cos\frac{\theta}{2}\end{array}\right).\end{equation}

The unnormalized post-measurement density matrices
$\tilde{\rho}_{0}$ and $\tilde{\rho}_{1}$ both have only one
non-zero eigenvalue respectively, i.e.,
$\lambda_{k}=\left[a^{2}+b^{2}+(-1)^{k}(a^{2}-b^{2})\cos\theta\right]/2$,
where $k=0,1$. That means the conditional information about the
lattice after the measurement of a local spin is zero, and
$\mathcal{J}$ is always equal to the entanglement of entropy $S$, no
matter what measurement we impose on the local spin. So the quantum
discord is equal to $S$,\begin{eqnarray}
D(\rho_{\textrm{GS}}) & = & \mathcal{J}=\mathcal{I}/2=S\nonumber \\
 & = & -a^{2}\log a^{2}-b^{2}\log b^{2},\end{eqnarray}
where\begin{eqnarray}
a^{2} & = & \left(1+\langle\theta_{0,0}\theta_{0,1}\rangle\right)/2,\nonumber \\
b^{2} & = & \left(1-\langle\theta_{0,0}\theta_{0,1}\rangle\right)/2.\end{eqnarray}

The quantum discord and mutual information of the global correlation
is shown in Fig. 3. Comparing with that of local spins correlation,
the quantum discord is not zero, which means the quantum correlation
exists in the lattice globally. It also signals the critical point
in the phase transition, just like the mutual information. As the
increase of $\beta$, the quantum discord decreases to zero, which
means the global quantum correlation disappears gradually .

In summary, the quantum correlation hides in the global lattice. We
can only get classical correlations between local spins. All these
results of correlations seem to suggest that the ground state of the
topologically ordered system behaves like a generalized GHZ state.
The quantum information is encoded in the lattice globally and so
it can be protected better than in other systems.

\section{Conclusion}

In this paper, we studied the correlations in Castelnovo-Chamon model.
Both local and global correlations were studied. The correlations
were measured by quantum discord. As we have seen, local spins are
classically correlated although the Hamiltonian is so complicated.
While the quantum correlation is hidden in the lattice globally. This
is quite peculiar comparing with previous studies. We analyzed that
these distinctive characters result from the high symmetry of the
2D topologically ordered system. The spins along any loop on the torus
are $Z_{2}$-symmetric. This strict constraint clears the quantum
correlations between local spins. Only global quantum correlation
exists, just like a generalized GHZ state. We believe that this is
a generic property in topological quantum phase transition because
of the particular symmetry of topologically order systems, as mentioned
previously.

Moreover, we calculate the mutual information of two nearest spins,
which signals critical behavior of the topological quantum phase transition.
Similar to previous study of fidelity, mutual information also works
as a local probe of the topologically ordered phase, although topological
order cannot be described by the symmetry-breaking of local order
parameter. More study is required for the correlations in other more
realistic systems.
\begin{acknowledgments}
The work is supported in part by the NSF of China Grant No. 90503009,
No. 10775116, and 973 Program Grant No. 2005CB724508.
\end{acknowledgments}

\end{document}